\documentstyle[12pt,axodraw]{article}
\newcommand{\NP}{Nucl. Phys. }
\newcommand{\PR}{Phys. Rev. }
\newcommand{\PRL}{Phys. Rev. Lett. }
\newcommand{\PL}{Phys. Lett. }

\begin{document}
\baselineskip=20pt

\pagenumbering{arabic}

\begin{flushright}
TU-HEP-TH-99/110
\end{flushright}

\vspace{1.0cm}
\begin{center}
{\Large\sf Associated production of Higgs bosons and heavy quarks at 
photon colliders}
\\[10pt]
\vspace{.5 cm}

{Jun-Yu Guo, Yi Liao}
\vspace{1.0ex}

{\small Department of Physics, Tsinghua University,
Beijing 100084, P.R.China\\}
\vspace{2.0ex}

{Yu-Ping Kuang}
\vspace{1.0ex}

{\small China Center of Advanced Science and Technology (World 
Laboratory), Beijing 100080, P.R.China\\}

\vspace{1.0ex}

{\small Department of Physics, Tsinghua University,
Beijing 100084, P.R.China\\}
\vspace{2.0ex}

{\bf Abstract}
\end{center}

The idea of fermion mass generation through Yukawa couplings between
Higgs bosons and fermions may be tested by their associated production
at future linear colliders ( LC ).
We study the production mechanism in the minimal supersymmeric standard
model at photon colliders that can be realized by the laser
back-scattering technique at the LC.
We find that the cross sections for the production of the light Higgs-top
pair, heavy Higgs-bottom pair and pseudoscalr Higgs-bottom pair can reach
the level of $\sim 1.0$ fb or higher for phenomenologically favoured values
of the parameters in the Higgs sector.
Since this mechanism is not affected by the appearance of Higgs resonances
as in electron collisions, it may provide a more sensitive way to determine
the Yukawa couplings between Higgs bosons and heavy quarks.

\begin{flushleft}


\end{flushleft}

\vspace{0.5cm}
\begin{center}
to appear in {\it Phys. Rev. D}
\end{center}

\baselineskip=24pt
\newpage

The Higgs particle that is assumed to trigger the spontaneous breakdown
of electroweak symmetries in the standard model (SM) or its extensions
has so far been experimentally evasive. The current negative result
of direct search imposes a lower bound on its mass, $m_H>100~$GeV
$\cite{search}$.
Yet, global fits of precision electroweak data favour a relatively light
SM Higgs boson with $m_H<260~$GeV at the $95\%$ confidence level
$\cite{fits}$.
In the minimal supersymmetric extension of the SM (MSSM), there are five
Higgs particles, the lightest of which cannot exceed a mass of about
$130~$GeV because of supersymmetry and upon including radiative
corrections$\cite{upper}$.
More generally, there are arguments from triviality and
vacuum stability that the SM Higgs boson mass should be in the narrow
range of $130~{\rm GeV}<m_H<180~{\rm GeV}$ if the scale of new physics
is of the order of the Planck scale$\cite{smrange}$
and that the lightest Higgs boson in general supersymmetric models
should be lighter than about $200~$GeV$\cite{susyrange}$.
All this implies a no-loss discovery of Higgs particles
at the current or proposed future high energy colliders if there exist
any such particles in Nature.

While the Higgs particles will probably be first discovered at LEP2,
Tevatron or the future LHC, the future linear collider (LC) will provide
an almost unique place to explore their detailed properties due to its
clean environment and high luminosity. The successful discovery of Higgs
particles will certainly advance our understanding of electroweak 
symmetry breaking and the mass generation of weak gauge bosons. However, 
it is important to notice that this does not mean that the issue of 
fermion mass generation will then automatically be solved since the 
latter could in principle be due to a separate ( or an additional )
origin. To answer this question, it will be necessary to examine the
Higgs-fermion interactions independently.

In the SM, fermions acquire mass through the Higgs mechanism and the
Yukawa couplings. Since these couplings are proportional to the mass of
fermions considered, only the Higgs-top coupling is phenomenologically
interesting. There are several ways to test the coupling. For example, 
we may measure the effective $H\gamma\gamma$ or $Hgg$ vertices ( $H$
for the SM Higgs boson, $\gamma$ the photon and $g$ the gluon ), which
essentially count the number of heavy particles, or the decay width of
$H\to t\bar{t}$ if this is kinematically allowed. If the Higgs boson
is not heavy, the Higgs boson radiation off top quarks at the future
LC will provide a promising mechanism to study the coupling
$\cite{smhiggs}\cite{susyhiggs}\cite{smqcd}$.
( For the same production mechanism at hadron colliders, see
Ref.$\cite{hadron}$. ) In the
MSSM, however, this simple pattern of associated production becomes
complicated$\cite{susyhiggs}\cite{susyqcd1}\cite{susyqcd2}$.
The Yukawa couplings are modified by the mixing angle
$\alpha$ in the CP-even sector of the neutral Higgs bosons and
$\tan\beta$, the ratio of the two vacuum expectation values of the
Higgs fields. It turns out that the top production will be suppressed
while the bottom enhanced as compared to the SM case for
phenomenologically favoured large values of $\tan\beta$. Furthermore, 
for associated Higgs-bottom production, the resonant structures of the 
Higgs bosons must be included due to the off-diagonal $A^0H^0Z$ and 
$A^0h^0Z$ interactions. ( Here $H^0$ and $h^0$ stand for the CP-even
heavy and light Higgs bosons, and $A^0$ the CP-odd one. ) The 
production rates are then overwhelmingly dominated by these
resonant structures, which make the extraction of Yukawa couplings very
involved and challenging. Actually, in the resonant region the
sensitivity to Yukawa couplings is largely reduced so that one
generally has to appeal to the continuum region for the extraction of
couplings$\cite{susyqcd2}\cite{communi}$.

In this work we shall consider an alternative way to measure the Yukawa
couplings at the LC; namely, the associated production in photon
collisions,
\begin{equation}
\gamma\gamma\to Q\bar{Q}\phi,~Q=t,~b,~\phi=h^0,~H^0,~A^0.
\end{equation}
The technique using back-scattering laser lights
$\cite{laser}$ makes this a possible
option for linear colliders which has been included in all proposed
projects for the future linear colliders. 
The advantages using photon-photon or photon-electron
collisions for some processes involving Higgs
production have been extensively explored in the literature
$\cite{gamma}\cite{egamma}$. For the
process at hand, the SM case has already been examined in Ref.
$\cite{smgamma}$. It was
found that the rate for associated Higgs-top production in photon
collisions is comparable to that in electron collisions while the
background is slightly better. In the supersymmetric case to be
investigated here, we generally cannot expect a rate as large as in
electron collisions because of the resonant enhancement mentioned above.
However, as far as the extraction of Yukawa couplings is concerned, one
has to subtract the resonant region so that the effective rate is much
smaller in electron collisions, while in photon collisions our processes
will not be contaminated by similar resonances.

Let us first specify the Yukawa interaction between the generic Higgs
boson $\phi$ and the fermion $\psi$,
\begin{equation}
\displaystyle
{\cal L}_Y=-\frac{m_f}{v}\bar{\psi}\left(a+ib\gamma_5\right)\psi\phi,
\end{equation}
where $m_f$ is the fermion mass and $v=2^{-1/4}G_F^{-1/2}$. Our
subsequent calculation applies to other models as well although we
are mainly interested in phenomenology in the SM and the MSSM. The
values for $a$ and $b$ are
\begin{equation}     
\begin{array}{rl}
\displaystyle
{\rm SM}:&\phi=H,~a=1,~b=0;\\
\\
\displaystyle
{\rm MSSM}:&\phi=h^0,~a=\left\{\begin{array}{ll}
       +\cos\alpha/\sin\beta,&{\rm for~up-type~fermions}\\
       -\sin\alpha/\cos\beta,&{\rm for~down-type~fermions}\\
      \end{array}\right.,~b=0,\\
\displaystyle
           &\phi=H^0,~a=\left\{\begin{array}{ll}
       +\sin\alpha/\sin\beta,&{\rm for~up-type~fermions}\\
       +\cos\alpha/\cos\beta,&{\rm for~down-type~fermions}\\
      \end{array}\right.,~b=0,\\
\displaystyle
           &\phi=A^0,~a=0,~b=\left\{\begin{array}{ll}
       -\cot\beta,&{\rm for~up-type~fermions}\\
       -\tan\beta,&{\rm for~down-type~fermions}\\
      \end{array}\right. .\\
\end{array}
\end{equation}
In the MSSM, we have four Higgs masses and two angles $\alpha$, $\beta$.
Supersymmetry imposes constraints on them so that only two of them are
independent, which we shall choose to be $m_{A^0}$ and $\tan\beta$.
These constraints are modified by radiative corrections.
For the numerical analysis to be presented later on, we have included the
contributions up to two loops in the effective potential approach
$\cite{carena}$.

The Feynman diagrams for the process $\gamma(k_1)\gamma(k_2)\to
f(p_1)\bar{f}(p_2)\phi(q)$ are shown in Fig. 1, from which we obtain
the amplitude,
\begin{equation}
\begin{array}{rcl}
\displaystyle
{\cal A}&=&\displaystyle
+e^2Q_f^2\frac{m_f}{v}\bar{u}(p_1){\cal A}_{\mu\nu}v(p_2)
\epsilon^{\mu}(1)\epsilon^{\nu}(2),\\
\\
\displaystyle
{\cal A}_{\mu\nu}&=&+Y(\rlap/p_1+\rlap/q+m_f)\gamma_{\mu}
(-2p_{2\nu}+\rlap/k_2\gamma_{\nu})(D_{2,2}D_1)^{-1}\\
\displaystyle
&&+(2p_{1\mu}-\gamma_{\mu}\rlap/k_1)Y
(-2p_{2\nu}+\rlap/k_2\gamma_{\nu})(D_{1,1}D_{2,2})^{-1}\\
\displaystyle
&&+(2p_{1\mu}-\gamma_{\mu}\rlap/k_1)\gamma_{\nu}(-\rlap/p_2-\rlap/q+m_f)Y
(D_{1,1}D_2)^{-1}\\
\displaystyle
&&+(k_1\leftrightarrow k_2,~\mu\leftrightarrow\nu),\\
\end{array}
\end{equation}
where $Q_f$ is the charge of the fermion, $\epsilon(i)(i=1,~2)$ are the
photon polarization vectors, and
\begin{equation}
\displaystyle Y=a+ib\gamma_5,~
D_{i,j}=-2k_i\cdot p_j,~D_i=m^2_{\phi}+2q\cdot p_i.\\
\end{equation}
The gauge invariance and the Bose symmetry with respect to the photons
are explicit in the above expression.

To obtain the total cross section $\sigma$, we should take the
convolution of the cross section $\hat{\sigma}$ for the subprocess
with the photon luminosity spectrum $F(x)$,
\begin{equation}
\displaystyle\sigma(s)=\int_{x_{\rm min}}^{x_{\rm max}}dx_1
\int_{x_{\rm min}x_{\rm max}/x_1}^{x_{\rm max}}dx_2
F(x_1)F(x_2)\hat{\sigma}(x_1x_2s),
\end{equation}
where $s$ is the c.m. energy squared for the parent electrons and the
subprocess occurs effectively at $\hat{s}=x_1x_2s$. Here, $x_i$ are
the fractions of the parent electrons' energies carried by the photons.
For photons produced in the laser back-scattering technique, we have,
\begin{equation}
\begin{array}{l}
\displaystyle F(x)=\frac{1}{D(\xi)}\left[1-x+\frac{1}{1-x}-
\frac{4x}{\xi(1-x)}+\frac{4x^2}{\xi^2(1-x)^2}\right],\\
\displaystyle D(\xi)=\left(1-\frac{4}{\xi}-\frac{8}{\xi^2}\right)
\ln(1+\xi)+\frac{1}{2}+\frac{8}{\xi}-\frac{1}{2(1+\xi)^2},\\
\displaystyle x_{\rm max}=\frac{\xi}{1+\xi},~
\xi=\frac{4E_0\omega_0}{m_e^2},\\
\end{array}
\end{equation}
where $E_0$ and $\omega_0$ are the incident electron and laser light
energies. To avoid unwanted $e^+e^-$ pair production from the collision
between the incident and back-scattered photons, we should not choose
too large $\omega_0$. This constrains the maximum value for $\xi$ to be
$\xi=2(1+\sqrt{2})$. The minimum value for $x$ is then determined by
the production threshold,
\begin{equation}
\displaystyle x_{\rm min}=\frac{\hat{s}_{\rm min}}{x_{\rm max}s},
~\hat{s}_{\rm min}=(2m_f+m_{\phi})^2.
\end{equation}

Our numerical results are shown in Figs. 2-5. The basic input parameters
are, $\alpha=1/128$, $m_W=80.33$ GeV, $m_Z=91.187$ GeV, $m_t=174$ GeV
and $m_b=5$ GeV. We choose $\mu=A_t=A_b=0$, and $\tilde{m}=1$ TeV
for the generic squark mass.
In the Higgs sector of the MSSM, we take $m_{A^0}$ and $\tan\beta$ as 
two independent parameters as mentioned above. To better understand our
results for production rates, let us first present in Fig. 2 the reduced
couplings squared $a^2_{h^0,~H^0}$ and the masses $m_{h^0,~H^0}$ of the
CP-even Higgs bosons as functions of $m_{A^0}$ at $\tan\beta=3$ and 
$30$. We observe the following features from the panel (a). First, for
fixed $\tan\beta$ and fermion type, the sum of $a^2_{h^0}$ and
$a^2_{H^0}$ is correspondingly fixed to be $\sin^{-2}\beta$ or
$\cos^{-2}\beta$ for up- or down-type fermions. Therefore, whenever the
reduced coupling decreases for the heavy Higgs boson
$H^0$, it increases for the light one $h^0$, and vice
versa. For large values of $\tan\beta$, the down-type fermion couplings
are enhanced while the up-type ones suppressed as compared to the SM
case. Second, the relative importance of the $H^0$ and $h^0$ couplings
to the same fermion is determined by $\tan\alpha$ which in turn depends
on $\tan\beta$ and $m_{A^0}$. For the parameters considered here,
$a^2_{h^0}>a^2_{H^0}$ for down-type fermions and $a^2_{h^0}<a^2_{H^0}$
for up-type fermions when $m_{A^0}$ is relatively small. The circumstance
is reversed as $m_{A^0}$ goes up. 
Third, in the limit of large $m_{A^0}$, $a^2_{h^0}$
approaches one for both type of fermions, and $a^2_{H^0}$ becomes
effectively the same as $b^2_{A^0}$. In this same limit, as shown in
the panel (b), $m_{H^0}$ also becomes the same as $m_{A^0}$ while
$m_{h^0}$ is saturated by its upper limit.

In Fig. 3 we plot the cross sections against the c.m. energy of the
subprocess for the $t\bar{t}\phi$ ( panel (a) ) and $b\bar{b}\phi$
( panel (b) ) production. It is clear that increasing the energy does
not gain much in cross sections once we are well above the production
threshold. This is especially true in the case of $b$ production, where
the rate actually decreases as the energy goes up from $500$ GeV. At
photon colliders using back-scattering laser lights the produced
photons are not monochromatic but constitute a continuous spectrum. This
effect is taken into account in Figs. 4 and 5 at two typical c.m.
energies of the parent electrons. The behaviour of cross sections
follows closely the pattern of the reduced couplings squared as we have
just discussed, up to modification factors introduced by the scattering
amplitude and the phase space. Let us first go over the top pair
production. 
When $\tan\beta$ is not too large, the associated $h^0$ production always
dominates, with a typical rate of about $1.0$ fb. 
The rate for $H^0$ depends on
the detail of the parameters $\tan\beta$ and $m_{A^0}$.
For small values of $m_{A^0}$, it is comparable to the rate for $h^0$
or even dominates when $\tan\beta$ is large enough.
The pseudoscalar $A^0$ is the most difficult
to produce. However, this hierarchy in cross sections is almost
reversed in the associated bottom-Higgs production which is
phenomenologically relevant only for large enough $\tan\beta$. 
Here we see that the production rate of the pseudoscalar $A^0$ is always
the largest and can reach several fb while the $h^0$ production is 
competitive only for relatively small $m_{A^0}$. The behaviour of $H^0$
in the large $m_{A^0}$ limit becomes identical to that of $A^0$ because
their couplings, masses and the scattering amplitudes squared approach
each other in this limit. Also shown in the panel (a)
of Figs. 4 and 5 is the SM model result for the $t\bar{t}H$ production.
It is clear that in the decoupling limit of large $m_{A^0}$ the $h^0$
production rate approaches correctly that of the SM Higgs $H$ with the
same mass as $h^0$, which is about $98$GeV for $\tan\beta=3$ and
$112$GeV for $\tan\beta=30$. Comparing Figs. 4 and 5
we realize that only the heavy $A^0$ and heavy $H^0$ production with
top pairs is considerably enhanced as the energy changes from 1 Tev 
to 2 TeV, which is consistent with the pattern observed in Figs. 2(a) 
and 3(a).

In summary, we have considered the associated production of Higgs
bosons and heavy quarks at the LC operating in the photon mode
which may be used to measure their Yukawa couplings.
For the SM case we reproduced
the previous result that the production rate is comparable to that
in the parent electron collisions. In the more complicated case of
the MSSM, the efficiency of extracting Yukawa couplings becomes somewhat
vague in the associated bottom-Higgs production of electron collisions
due to the contamination by the Higgs resonances. Instead, we proposed
to determine these couplings through photon collisions where no such
resonances are involved. We found that the production rate for
$t\bar{t}h^0$ is as large as $1.0$ fb while
$b\bar{b}H^0$ and $b\bar{b}A^0$ can be equally or more copiously
produced for phenomenologically favoured large values of $\tan\beta$.
This amounts to about $500$ events or more for an annually
integrated luminosity of $500{\rm fb}^{-1}$ at the TESLA.
In this manner, the measurement at photon colliders may provide us
a global picture of the Yukawa couplings between the neutral Higgs
bosons and the heavy quarks. We also found that a high
luminosity is more essential than a high c.m. energy for this purpose.
Finally, we should mention that we have considered only the lowest
order contribution which may receive potentially large radiative
corrections especially for the bottom pair production in the case of
large $\tan\beta$.

{\bf Acknowledgements}. Y.L. would like to thank M. Spira and P. M. Zerwas
for helpful discussions concerning the effect of Higgs resonances on the
associated bottom-Higgs production in electron collisions. He is also
grateful to M. Spira for providing us the Fortran code$\cite{fcode}$ to
compute the Higgs masses and couplings based on the work in
Ref.$\cite{carena}$.

%\baselineskip=20pt
%\newpage

%\newpage
\begin{flushleft}
{\Large Figure Captions }
\end{flushleft}
\noindent
Fig. 1 Feynman diagrams for the subprocess
$\gamma\gamma\to f\bar{f}\phi$. Crossing diagrams are not shown.

\noindent
Fig. 2 Reduced couplings squared $a^2_{h^0,~H^0}$ ( panel (a) ) and 
masses $m_{h^0,~H^0}$ ( panel (b) ) as functions of $m_{A^0}$ for
$\tan\beta=3$ ( solid ) and $30$ ( dashed ). $b^2_{A^0}$ is fixed
to be $\tan^2\beta$ or $\cot^2\beta$ which is not shown.

\noindent
Fig. 3 Cross sections as functions of the c.m. energy of monochromatic
photons for $m_{A^0}=150$ GeV, $\tan\beta=3$ ( solid ) and $30$ 
( dashed ): (a) $t\bar{t}\phi$ production; (b) $b\bar{b}\phi$
production.

\noindent
Fig. 4 Cross sections as functions of $m_{A^0}$ in the laser
back-scattering technique at the electrons' energy $\sqrt{s}=1$ TeV
for $\tan\beta=3$ ( solid ) and $30$ ( dashed ):
(a) $t\bar{t}\phi$ production; (b) $b\bar{b}\phi$ production.
The long-dashed curve in the panel (a) stands for the SM Higgs for
which $m_{A^0}$ should be understood as $m_H$.

\noindent
Fig. 5 Same as Fig. 4, but at $\sqrt{s}=2$ TeV.

%%%%%%%%%%%%%%%%%%%%%%%%%%%%%%%%%%%%%%%%%%%%%%%%
%%%%%%%%%     Drawing Figure 1
%\newpage
\begin{center}
\begin{picture}(300,100)(0,0)
\SetWidth{1.}

\Photon(10,25)(40,25){3}{4}
\Photon(10,75)(40,75){3}{4}
\Line(90,25)(40,25)
\ArrowLine(40,25)(40,75)
\Line(40,75)(90,75)
\DashLine(60,75)(90,50){3.}

\SetOffset(100,0)

\Photon(10,25)(40,25){3}{4}
\Photon(10,75)(40,75){3}{4}
\Line(90,25)(40,25)
\ArrowLine(40,25)(40,75)
\Line(40,75)(90,75)
\DashLine(40,50)(90,50){3.}
\Text(50,0)[]{Figure $1$}

\SetOffset(200,0)

\Photon(10,25)(40,25){3}{4}
\Photon(10,75)(40,75){3}{4}
\Line(90,25)(40,25)
\ArrowLine(40,25)(40,75)
\Line(40,75)(90,75)
\DashLine(60,25)(90,50){3.}

\end{picture}\\
\end{center}
%%%%%%%%%%%%%%%%%%%%%%%%%%%%%%%%%%%%%%%%%%%%%%%%
%%%%%%%%%     Drawing Figure 2
\newpage
\begin{center}
\begin{picture}(350,550)(0,0)

%Panel (a)
\SetOffset(40,300)\SetWidth{1.}
\LinAxis(0,0)(300,0)(4,5,5,0,1.5)
\LinAxis(0,200)(300,200)(4,5,-5,0,1.5)
\LogAxis(0,0)(0,200)(6,-5,0,1.5)
\LogAxis(300,0)(300,200)(6,5,0,1.5)
\Text(0,-10)[]{$100$}
\Text(75,-10)[]{$150$}
\Text(150,-10)[]{$200$}
\Text(225,-10)[]{$250$}
\Text(300,-10)[]{$300$}
\Text(140,-25)[]{$m_{A^0}$(GeV)}
\Text(-15,0)[]{$10^{-3}$}
\Text(-15,33)[]{$10^{-2}$}
\Text(-15,67)[]{$10^{-1}$}
\Text(-15,100)[]{$10^0~~$}
\Text(-15,133)[]{$10^{+1}$}
\Text(-15,167)[]{$10^{+2}$}
\Text(-15,200)[]{$10^{+3}$}
\Text(-40,100)[]{$a^2_{h^0,H^0}$}
\Text(285,185)[]{$(a)$}
\Text(260,95)[]{$h^0tt$}
\Curve{(0.00E+00,85.96313)
(15.00000,89.02716)(30.00000,91.43711)
(45.00000,93.26833)(60.00000,94.63754)
(75.00000,95.65987)(90.00000,96.42951)
(105.0000,97.01672)(120.0000,97.47366)
(135.0000,97.83499)(150.0000,98.12439)
(165.0000,98.35957)(180.0000,98.55323)
(195.0000,98.71467)(210.0000,98.85073)
(225.0000,98.96655)(240.0000,99.06603)
(255.0000,99.15218)(270.0000,99.22733)
(285.0000,99.29333)(300.0000,99.35167)
}
\Text(20,135)[]{$h^0bb$}
\Curve{(0.00E+00,127.2896)
(15.00000,125.4038)(30.00000,123.3526)
(45.00000,121.2724)(60.00000,119.2728)
(75.00000,117.4206)(90.00000,115.7449)
(105.0000,114.2499)(120.0000,112.9090)
(135.0000,111.7133)(150.0000,110.6566)
(165.0000,109.7215)(180.0000,108.8922)
(195.0000,108.1547)(210.0000,107.4967)
(225.0000,106.9077)(240.0000,106.3787)
(255.0000,105.9020)(270.0000,105.4712)
(285.0000,105.0805)(300.0000,104.7252)
}
\Text(260,70)[]{$H^0tt$}
\SetWidth{2.}
\Curve{(0.00E+00,95.48150)
(15.00000,93.59572)(30.00000,91.54452)
(45.00000,89.46432)(60.00000,87.46475)
(75.00000,85.61253)(90.00000,83.93684)
(105.0000,82.44184)(120.0000,81.11980)
(135.0000,79.95416)(150.0000,78.92601)
(165.0000,78.01814)(180.0000,77.21481)
(195.0000,76.50209)(210.0000,75.86785)
(225.0000,75.30167)(240.0000,74.79465)
(255.0000,74.33916)(270.0000,73.92870)
(285.0000,73.55770)(300.0000,73.22142)
}
\SetWidth{1.}
\Text(260,140)[]{$H^0bb$}
\SetWidth{2.}
\Curve{(0.00E+00,117.7712)
(15.00000,120.8353)(30.00000,123.2452)
(45.00000,125.0764)(60.00000,126.4456)
(75.00000,127.4680)(90.00000,128.2376)
(105.0000,128.8248)(120.0000,129.2629)
(135.0000,129.5941)(150.0000,129.8550)
(165.0000,130.0629)(180.0000,130.2307)
(195.0000,130.3673)(210.0000,130.4796)
(225.0000,130.5725)(240.0000,130.6501)
(255.0000,130.7150)(270.0000,130.7698)
(285.0000,130.8161)(300.0000,130.8554)
}
\SetWidth{1.}
\Text(60,106)[]{$h^0tt$}
\DashCurve{(0.00E+00,54.46627)
(15.00000,82.83688)(30.00000,98.19315)
(45.00000,99.52029)(60.00000,99.77727)
(75.00000,99.86831)(90.00000,99.91138)
(105.0000,99.93542)(120.0000,99.95039)
(135.0000,99.96041)(150.0000,99.96752)
(165.0000,99.97275)(180.0000,99.97675)
(195.0000,99.97988)(210.0000,99.98239)
(225.0000,99.98443)(240.0000,99.98613)
(255.0000,99.98755)(270.0000,99.98876)
(285.0000,99.98979)(300.0000,99.99069)
}{5}
\Text(60,150)[]{$h^0bb$}
\DashCurve{(0.00E+00,197.8545)
(15.00000,193.2189)(30.00000,167.5925)
(45.00000,149.3980)(60.00000,138.9513)
(75.00000,132.0472)(90.00000,127.0805)
(105.0000,123.3145)(120.0000,120.3428)
(135.0000,117.9366)(150.0000,115.9545)
(165.0000,114.2958)(180.0000,112.8896)
(195.0000,111.6845)(210.0000,110.6419)
(225.0000,109.7328)(240.0000,108.9342)
(255.0000,108.2285)(270.0000,107.6011)
(285.0000,107.0407)(300.0000,106.5375)
}{5}
\Text(260,20)[]{$H^0tt$}
\SetWidth{2.}
\DashCurve{(0.00E+00,99.37975)
(15.00000,94.74416)(30.00000,69.11769)
(45.00000,50.92323)(60.00000,40.47659)
(75.00000,33.57243)(90.00000,28.60579)
(105.0000,24.83972)(120.0000,21.88012)
(135.0000,19.49310)(150.0000,17.52928)
(165.0000,15.88794)(180.0000,14.49830)
(195.0000,13.30896)(210.0000,12.28158)
(225.0000,11.38697)(240.0000,10.60245)
(255.0000,9.910149)(270.0000,9.295794)
(285.0000,8.747841)(300.0000,8.256854)
}{5}
\SetWidth{1.}
\Text(260,190)[]{$H^0bb$}
\SetWidth{2.}
\DashCurve{(0.00E+00,152.9410)
(15.00000,181.3116)(30.00000,196.6679)
(45.00000,197.9950)(60.00000,198.2520)
(75.00000,198.3431)(90.00000,198.3861)
(105.0000,198.4102)(120.0000,198.4131)
(135.0000,198.4039)(150.0000,198.3928)
(165.0000,198.3806)(180.0000,198.3681)
(195.0000,198.3554)(210.0000,198.3427)
(225.0000,198.3302)(240.0000,198.3179)
(255.0000,198.3059)(270.0000,198.2941)
(285.0000,198.2826)(300.0000,198.2714)
}{5}

%Panel (b)
\SetOffset(40,50)\SetWidth{1.}
\LinAxis(0,0)(300,0)(4,5,5,0,1.5)
\LinAxis(0,200)(300,200)(4,5,-5,0,1.5)
\LinAxis(0,200)(0,0)(6,5,5,0,1.5)
\LinAxis(300,0)(300,200)(6,5,5,0,1.5)
\Text(0,-10)[]{$100$}
\Text(75,-10)[]{$150$}
\Text(150,-10)[]{$200$}
\Text(225,-10)[]{$250$}
\Text(300,-10)[]{$300$}
\Text(140,-25)[]{$m_{A^0}$(GeV)}
\Text(-15,0)[]{$50$}
\Text(-15,33)[]{$100$}
\Text(-15,67)[]{$150$}
\Text(-15,100)[]{$200$}
\Text(-15,133)[]{$250$}
\Text(-15,167)[]{$300$}
\Text(-15,200)[]{$350$}
\Text(-40,100)[]{$m_{h^0,H^0}$}
\Text(-40,90)[]{(GeV)}
\Text(285,185)[]{$(b)$}
\Text(140,-50)[]{Figure $2$}
\Text(225,20)[]{$h^0$}
\Curve{(0.00E+00,18.28796)
(15.00000,20.95836)(30.00000,23.10308)
(45.00000,24.79476)(60.00000,26.12029)
(75.00000,27.16125)(90.00000,27.98512)
(105.0000,28.64424)(120.0000,29.17782)
(135.0000,29.61494)(150.0000,29.97715)
(165.0000,30.28049)(180.0000,30.53703)
(195.0000,30.75594)(210.0000,30.94427)
(225.0000,31.10751)(240.0000,31.24996)
(255.0000,31.37505)(270.0000,31.48550)
(285.0000,31.58355)(300.0000,31.67101)
}
\Text(225,150)[]{$H^0$}
\Curve{(0.00E+00,54.20414)
(15.00000,57.84404)(30.00000,62.05040)
(45.00000,66.74343)(60.00000,71.83083)
(75.00000,77.22674)(90.00000,82.86025)
(105.0000,88.67629)(120.0000,94.63338)
(135.0000,100.7006)(150.0000,106.8547)
(165.0000,113.0784)(180.0000,119.3586)
(195.0000,125.6849)(210.0000,132.0496)
(225.0000,138.4464)(240.0000,144.8704)
(255.0000,151.3176)(270.0000,157.7847)
(285.0000,164.2690)(300.0000,170.7685)
}
\Text(225,50)[]{$h^0$}
\DashCurve{(0.00E+00,32.91041)
(15.00000,38.65255)(30.00000,40.86273)
(45.00000,41.21930)(60.00000,41.33432)
(75.00000,41.38961)(90.00000,41.42178)
(105.0000,41.44269)(120.0000,41.45729)
(135.0000,41.46801)(150.0000,41.47619)
(165.0000,41.48261)(180.0000,41.48776)
(195.0000,41.49199)(210.0000,41.49549)
(225.0000,41.49844)(240.0000,41.50096)
(255.0000,41.50312)(270.0000,41.50500)
(285.0000,41.50664)(300.0000,41.50808)
}{5}
\Text(225,120)[]{$H^0$}
\DashCurve{(0.00E+00,42.01956)
(15.00000,42.94024)(30.00000,47.39331)
(45.00000,53.70036)(60.00000,60.24926)
(75.00000,66.85813)(90.00000,73.49035)
(105.0000,80.13401)(120.0000,86.78416)
(135.0000,93.43832)(150.0000,100.0952)
(165.0000,106.7539)(180.0000,113.4139)
(195.0000,120.0750)(210.0000,126.7369)
(225.0000,133.3995)(240.0000,140.0625)
(255.0000,146.7260)(270.0000,153.3898)
(285.0000,160.0538)(300.0000,166.7182)
}{5}

\end{picture}\\
\end{center}

%%%%%%%%%%%%%%%%%%%%%%%%%%%%%%%%%%%%%%%%%%%%%%%%
%%%%%%%%%     Drawing Figure 3
\newpage
\begin{center}
\begin{picture}(350,550)(0,0)

%Panel (a)
\SetOffset(40,300)\SetWidth{1.}
\LinAxis(0,0)(300,0)(3,5,5,0,1.5)
\LinAxis(0,200)(300,200)(3,5,-5,0,1.5)
\LogAxis(0,0)(0,200)(5,-5,0,1.5)
\LogAxis(300,0)(300,200)(5,5,0,1.5)
\Text(0,-10)[]{$500$}
\Text(100,-10)[]{$1000$}
\Text(200,-10)[]{$1500$}
\Text(300,-10)[]{$2000$}
\Text(140,-25)[]{$\sqrt{s_{\gamma\gamma}}$(GeV)}
\Text(-15,0)[]{$10^{-4}$}
\Text(-15,40)[]{$10^{-3}$}
\Text(-15,80)[]{$10^{-2}$}
\Text(-15,120)[]{$10^{-1}$}
\Text(-15,160)[]{$10^{0}~~$}
\Text(-15,200)[]{$10^{+1}$}
\Text(-40,100)[]{$\sigma$(fb)}
\Text(285,185)[]{$(a)$}
\Text(20,185)[]{$h^0$}
\Curve{(20.00000,177.9304)
(40.00000,181.0450)(60.00000,181.0289)
(80.00000,180.1044)(100.0000,178.7883)
(120.0000,177.5447)(140.0000,176.1967)
(160.0000,175.0013)(180.0000,173.9397)
(200.0000,173.0189)(220.0000,172.1735)
(240.0000,171.2592)(260.0000,170.5235)
(280.0000,169.6737)(300.0000,169.0602)
}
\Text(20,165)[]{$h^0$}
\DashCurve{(20.00000,172.5940)
(40.00000,177.9794)(60.00000,179.0413)
(80.00000,178.8373)(100.0000,178.0607)
(120.0000,177.2771)(140.0000,176.3093)
(160.0000,175.4141)(180.0000,174.6603)
(200.0000,173.9886)(220.0000,173.3909)
(240.0000,172.6652)(260.0000,172.0928)
(280.0000,171.4389)(300.0000,170.9506)
}{5}
\Text(20,135)[]{$H^0$}
\Curve{(20.00000,125.1275)
(40.00000,139.6572)(60.00000,144.4849)
(80.00000,146.6728)(100.0000,147.6410)
(120.0000,148.2768)(140.0000,148.4556)
(160.0000,148.4776)(180.0000,148.5853)
(200.0000,148.5853)(220.0000,148.5818)
(240.0000,148.3980)(260.0000,148.2554)
(280.0000,148.0200)(300.0000,147.8204)
}
\Text(20,65)[]{$H^0$}
\DashCurve{(20.00000,72.42944)
(40.00000,83.58286)(60.00000,87.12564)
(80.00000,88.52294)(100.0000,88.92170)
(120.0000,89.10610)(140.0000,88.92578)
(160.0000,88.65911)(180.0000,88.51126)
(200.0000,88.31049)(220.0000,88.13895)
(240.0000,87.79259)(260.0000,87.52358)
(280.0000,87.17815)(300.0000,86.89404)
}{5}
\Text(20,95)[]{$A^0$}
\Curve{(20.00000,83.45225)
(40.00000,105.0977)(60.00000,115.2201)
(80.00000,120.8416)(100.0000,124.3828)
(120.0000,126.6499)(140.0000,128.0568)
(160.0000,128.7606)(180.0000,129.5502)
(200.0000,129.7589)(220.0000,130.0179)
(240.0000,129.9926)(260.0000,130.0904)
(280.0000,129.8151)(300.0000,129.6949)
}
\Text(20,15)[]{$A^0$}
\DashCurve{(20.00000,3.452247)
(40.00000,25.09768)(60.00000,35.22008)
(80.00000,40.84158)(100.0000,44.38280)
(120.0000,46.64993)(140.0000,48.05677)
(160.0000,48.76057)(180.0000,49.55020)
(200.0000,49.75888)(220.0000,50.01789)
(240.0000,49.99258)(260.0000,50.09039)
(280.0000,49.81509)(300.0000,49.69493)
}{5}

%Panel (b)
\SetOffset(40,50)\SetWidth{1.}
\LinAxis(0,0)(300,0)(3,5,5,0,1.5)
\LinAxis(0,200)(300,200)(3,5,-5,0,1.5)
\LogAxis(0,0)(0,200)(5,-5,0,1.5)
\LogAxis(300,0)(300,200)(5,5,0,1.5)
\Text(0,-10)[]{$500$}
\Text(100,-10)[]{$1000$}
\Text(200,-10)[]{$1500$}
\Text(300,-10)[]{$2000$}
\Text(140,-25)[]{$\sqrt{s_{\gamma\gamma}}$(GeV)}
\Text(-15,0)[]{$10^{-4}$}
\Text(-15,40)[]{$10^{-3}$}
\Text(-15,80)[]{$10^{-2}$}
\Text(-15,120)[]{$10^{-1}$}
\Text(-15,160)[]{$10^{0}~~$}
\Text(-15,200)[]{$10^{+1}$}
\Text(-40,100)[]{$\sigma$(fb)}
\Text(285,185)[]{$(b)$}
\Text(140,-50)[]{Figure $3$}
\Text(20,85)[]{\small{$h^0$}}
\Curve{(0.00E+00,94.39348)
(20.00000,92.09023)(40.00000,89.69834)
(60.00000,87.92223)(80.00000,86.23392)
(100.0000,84.44487)(120.0000,82.00970)
(140.0000,81.11990)(160.0000,79.26921)
(180.0000,78.10040)(200.0000,75.67182)
(220.0000,74.99088)(240.0000,74.02564)
(260.0000,72.33167)(280.0000,71.55808)
(300.0000,70.15843)
}
\Text(20,115)[]{\small{$h^0$}}
\DashCurve{(0.00E+00,109.2549)
(20.00000,107.3765)(40.00000,105.2954)
(60.00000,103.7047)(80.00000,101.9279)
(100.0000,100.1028)(120.0000,98.16202)
(140.0000,96.59868)(160.0000,95.65141)
(180.0000,94.74472)(200.0000,92.05186)
(220.0000,92.13615)(240.0000,89.92500)
(260.0000,88.32465)(280.0000,87.99424)
(300.0000,86.40085)
}{5}
\Text(200,80)[]{\small{$H^0$}}
\Curve{(0.00E+00,96.99983)
(20.00000,96.23177)(40.00000,94.94960)
(60.00000,93.60472)(80.00000,91.48640)
(100.0000,90.97780)(120.0000,88.95347)
(140.0000,87.68795)(160.0000,87.04489)
(180.0000,85.73068)(200.0000,83.28969)
(220.0000,84.13601)(240.0000,81.61660)
(260.0000,80.68991)(280.0000,80.35633)
(300.0000,78.31319)
}
\Text(20,175)[]{\small{$H^0$}}
\DashCurve{(0.00E+00,183.8375)
(20.00000,182.9165)(40.00000,181.5479)
(60.00000,179.9065)(80.00000,178.8198)
(100.0000,177.1030)(120.0000,174.8227)
(140.0000,173.7573)(160.0000,172.9257)
(180.0000,171.8826)(200.0000,169.1654)
(220.0000,169.6176)(240.0000,167.2485)
(260.0000,166.4309)(280.0000,166.0991)
(300.0000,164.1724)
}{5}
\Text(50,100)[]{\small{$A^0$}}
\Curve{(0.00E+00,104.1295)
(20.00000,103.2004)(40.00000,101.5497)
(60.00000,100.3379)(80.00000,98.66222)
(100.0000,97.23219)(120.0000,95.26003)
(140.0000,93.45828)(160.0000,92.47393)
(180.0000,92.02278)(200.0000,89.18542)
(220.0000,89.06915)(240.0000,87.38865)
(260.0000,86.75797)(280.0000,86.05840)
(300.0000,85.33126)
}
\Text(20,190)[]{\small{$A^0$}}
\DashCurve{(0.00E+00,184.1295)
(20.00000,183.2004)(40.00000,181.5497)
(60.00000,180.3379)(80.00000,178.6622)
(100.0000,177.2322)(120.0000,175.2600)
(140.0000,173.4583)(160.0000,172.4739)
(180.0000,172.0228)(200.0000,169.1854)
(220.0000,169.0691)(240.0000,167.3887)
(260.0000,166.7580)(280.0000,166.0584)
(300.0000,165.3313)
}{5}

\end{picture}\\
\end{center}

%%%%%%%%%%%%%%%%%%%%%%%%%%%%%%%%%%%%%%%%%%%%%%%%
%%%%%%%%%     Drawing Figure 4
\newpage
\begin{center}
\begin{picture}(350,550)(0,0)

%Panel (a)
\SetOffset(40,300)\SetWidth{1.}
\LinAxis(0,0)(300,0)(4,5,5,0,1.5)
\LinAxis(0,200)(300,200)(4,5,-5,0,1.5)
\LogAxis(0,0)(0,200)(5,-5,0,1.5)
\LogAxis(300,0)(300,200)(5,5,0,1.5)
\Text(0,-10)[]{$100$}
\Text(75,-10)[]{$150$}
\Text(150,-10)[]{$200$}
\Text(225,-10)[]{$250$}
\Text(300,-10)[]{$300$}
\Text(140,-25)[]{$m_{A^0}$(GeV)}
\Text(-15,0)[]{$10^{-4}$}
\Text(-15,40)[]{$10^{-3}$}
\Text(-15,80)[]{$10^{-2}$}
\Text(-15,120)[]{$10^{-1}$}
\Text(-15,160)[]{$10^{0}~~$}
\Text(-15,200)[]{$10^{+1}$}
\Text(-40,100)[]{$\sigma$(fb)}
\Text(285,185)[]{$(a)$}
\Text(150,170)[]{$h^0$}
\Curve{(0.00E+00,156.6234)
(15.00000,158.3622)(30.00000,159.7175)
(45.00000,160.7264)(60.00000,161.4423)
(75.00000,161.9199)(90.00000,162.2553)
(105.0000,162.5265)(120.0000,162.6872)
(135.0000,162.8213)(150.0000,162.9299)
(165.0000,163.0235)(180.0000,163.0254)
(195.0000,163.0708)(210.0000,163.1536)
(225.0000,163.1899)(240.0000,163.1655)
(255.0000,163.1905)(270.0000,163.1952)
(285.0000,163.1947)(300.0000,163.2236)
}
\Text(150,150)[]{$h^0$}
\DashCurve{(0.00E+00,108.4990)(15.00000,138.5791)
(30.00000,155.4915)(45.00000,156.8549)
(60.00000,157.0951)(75.00000,157.1481)
(90.00000,157.1690)(105.0000,157.2096)
(120.0000,157.2039)(135.0000,157.2163)
(150.0000,157.2305)(165.0000,157.2540)
(180.0000,157.2038)(195.0000,157.2080)
(210.0000,157.2607)(225.0000,157.2729)
(240.0000,157.2256)(255.0000,157.2361)
(270.0000,157.2262)(285.0000,157.2156)
(300.0000,157.2350)
}{5}
\Text(150,95)[]{$H^0$}
\Curve{(0.00E+00,143.2154)(15.00000,138.4765)
(30.00000,133.1603)(45.00000,127.5035)
(60.00000,121.6564)(75.00000,115.7230)
(90.00000,109.8233)(105.0000,103.9953)
(120.0000,98.20264)(135.0000,92.48922)
(150.0000,86.77734)(165.0000,81.09773)
(180.0000,75.36137)(195.0000,69.62058)
(210.0000,63.87955)(225.0000,58.07803)
(240.0000,52.13562)(255.0000,46.13715)
(270.0000,40.05063)(285.0000,33.91145)
(300.0000,27.67532)
}
\Text(150,30)[]{$H^0$}
\DashCurve{(0.00E+00,156.1492)
(15.00000,149.9429)(30.00000,116.1615)
(45.00000,90.07714)(60.00000,73.12741)
(75.00000,60.34770)(90.00000,49.86782)
(105.0000,40.80785)(120.0000,32.63638)
(135.0000,25.11889)(150.0000,18.00916)
(165.0000,11.21965)(180.0000,4.588290)
%(195.0000,-1.884949)(210.0000,-8.228724)
%(225.0000,-14.54010)(240.0000,-20.91752)
%(255.0000,-27.28578)(270.0000,-33.69918)
%(285.0000,-40.13661)(300.0000,-46.64392)
}{5}
\Text(150,77)[]{$A^0$}
\Curve{(0.00E+00,92.54755)
(15.00000,90.41171)(30.00000,88.27869)
(45.00000,86.09160)(60.00000,83.90139)
(75.00000,81.70121)(90.00000,79.49811)
(105.0000,77.26087)(120.0000,74.99520)
(135.0000,72.74304)(150.0000,70.49359)
(165.0000,68.15730)(180.0000,65.71056)
(195.0000,63.32511)(210.0000,60.89724)
(225.0000,58.31609)(240.0000,55.66012)
(255.0000,52.95422)(270.0000,50.15064)
(285.0000,47.23332)(300.0000,44.17261)
}
\Text(40,15)[]{$A^0$}
\DashCurve{(0.00E+00,12.54755)
(15.00000,10.41172)(30.00000,8.278687)
(45.00000,6.091597)(60.00000,3.901390)
(75.00000,1.701207)%(90.00000,-0.5018896)
%(105.0000,-2.739128)(120.0000,-5.012970)
%(135.0000,-7.278172)(150.0000,-9.540019)
%(165.0000,-11.88813)(180.0000,-14.34614)
%(195.0000,-16.74239)(210.0000,-19.18060)
%(225.0000,-21.77169)(240.0000,-24.43722)
%(255.0000,-27.15234)(270.0000,-29.96481)
%(285.0000,-32.89070)(300.0000,-35.95972)
}{5}
%SM Higgs
\Text(150,125)[]{${\rm SM}~H$}
\DashCurve{(0.00E+00,162.8777)(15.00000,158.4175)
(21.60000,156.2910)(30.00000,153.7383)
(43.34999,149.7747)(45.00000,149.3035)
(60.00000,144.7271)(75.00000,140.3750)
(90.00000,135.6696)(105.0000,131.0224)
(120.0000,126.3646)(135.0000,121.6926)
(150.0000,116.9991)(165.0000,112.1341)
(180.0000,107.2089)(195.0000,102.1790)
(210.0000,97.04800)(225.0000,91.81881)
(240.0000,86.43307)(255.0000,80.80865)
(270.0000,75.25822)(285.0000,69.48248)
(300.0000,63.34576)
}{12}

%Panel (b)
\SetOffset(40,50)\SetWidth{1.}
\LinAxis(0,0)(300,0)(4,5,5,0,1.5)
\LinAxis(0,200)(300,200)(4,5,-5,0,1.5)
\LogAxis(0,0)(0,200)(5,-5,0,1.5)
\LogAxis(300,0)(300,200)(5,5,0,1.5)
\Text(0,-10)[]{$100$}
\Text(75,-10)[]{$150$}
\Text(150,-10)[]{$200$}
\Text(225,-10)[]{$250$}
\Text(300,-10)[]{$300$}
\Text(140,-25)[]{$m_{A^0}$(GeV)}
\Text(-15,0)[]{$10^{-4}$}
\Text(-15,40)[]{$10^{-3}$}
\Text(-15,80)[]{$10^{-2}$}
\Text(-15,120)[]{$10^{-1}$}
\Text(-15,160)[]{$10^{0}~~$}
\Text(-15,200)[]{$10^{+1}$}
\Text(-40,100)[]{$\sigma$(fb)}
\Text(285,185)[]{$(b)$}
\Text(140,-50)[]{Figure $4$}
\Text(15,100)[]{$h^0$}
\Curve{(0.00E+00,107.7089)
(15.00000,104.5699)(30.00000,101.3956)
(45.00000,98.39793)(60.00000,95.58370)
(75.00000,93.06609)(90.00000,90.88483)
(105.0000,88.85169)(120.0000,87.00970)
(135.0000,85.56326)(150.0000,84.28045)
(165.0000,82.96742)(180.0000,81.91438)
(195.0000,80.95856)(210.0000,80.17026)
(225.0000,79.30410)(240.0000,78.68701)
(255.0000,78.01045)(270.0000,77.54421)
(285.0000,77.02629)(300.0000,76.47890)
}
\Text(80,115)[]{$h^0$}
\DashCurve{(0.00E+00,188.0313)
(15.00000,180.7625)(30.00000,149.4924)
(45.00000,127.4072)(60.00000,114.9617)
(75.00000,106.6346)(90.00000,100.6914)
(105.0000,96.05466)(120.0000,92.45515)
(135.0000,89.71642)(150.0000,87.41283)
(165.0000,85.32578)(180.0000,83.63204)
(195.0000,82.23479)(210.0000,80.97771)
(225.0000,79.76435)(240.0000,78.87547)
(255.0000,77.95327)(270.0000,77.30173)
(285.0000,76.61780)(300.0000,75.92817)
}{8}
\Text(15,83)[]{$H^0$}
\Curve{(0.00E+00,86.27166)
(15.00000,88.78910)(30.00000,90.62679)
(45.00000,91.60118)(60.00000,92.20449)
(75.00000,91.97930)(90.00000,91.63146)
(105.0000,90.87041)(120.0000,89.83331)
(135.0000,89.24464)(150.0000,88.09692)
(165.0000,86.89877)(180.0000,85.70760)
(195.0000,84.56077)(210.0000,83.29152)
(225.0000,81.98923)(240.0000,80.80843)
(255.0000,79.39682)(270.0000,78.18571)
(285.0000,76.73983)(300.0000,75.55945)
}
\Text(15,150)[]{$H^0$}
\DashCurve{(0.00E+00,131.6396)
(15.00000,165.3092)(30.00000,182.5732)
(45.00000,182.3345)(60.00000,181.2047)
(75.00000,179.5763)(90.00000,178.0241)
(105.0000,176.3688)(120.0000,174.6425)
(135.0000,173.4887)(150.0000,171.8792)
(165.0000,170.2815)(180.0000,168.7672)
(195.0000,167.3810)(210.0000,165.8805)
(225.0000,164.3797)(240.0000,163.0291)
(255.0000,161.4791)(270.0000,160.1334)
(285.0000,158.5846)(300.0000,157.2984)
}{5}
\Text(15,115)[]{$A^0$}
\Curve{(0.00E+00,108.8175)
(15.00000,106.9614)(30.00000,105.0521)
(45.00000,103.2797)(60.00000,101.7958)
(75.00000,99.88647)(90.00000,98.40978)
(105.0000,96.66830)(120.0000,95.01820)
(135.0000,93.55252)(150.0000,91.84718)
(165.0000,90.35023)(180.0000,88.75777)
(195.0000,87.47507)(210.0000,85.86295)
(225.0000,84.36583)(240.0000,82.93211)
(255.0000,81.47024)(270.0000,80.13483)
(285.0000,78.58482)(300.0000,77.21586)
}
\Text(80,190)[]{$A^0$}
\DashCurve{(0.00E+00,188.8175)
(15.00000,186.9614)(30.00000,185.0521)
(45.00000,183.2797)(60.00000,181.7958)
(75.00000,179.8865)(90.00000,178.4098)
(105.0000,176.6683)(120.0000,175.0264)
(135.0000,173.5737)(150.0000,171.8808)
(165.0000,170.3956)(180.0000,168.8145)
(195.0000,167.5426)(210.0000,165.9408)
(225.0000,164.4536)(240.0000,163.0294)
(255.0000,161.5768)(270.0000,160.2503)
(285.0000,158.7088)(300.0000,157.3482)
}{4}

\end{picture}\\
\end{center}

%%%%%%%%%%%%%%%%%%%%%%%%%%%%%%%%%%%%%%%%%%%%%%%%
%%%%%%%%%     Drawing Figure 5
\newpage
\begin{center}
\begin{picture}(350,550)(0,0)

%Panel (a)
\SetOffset(40,300)\SetWidth{1.}
\LinAxis(0,0)(300,0)(4,5,5,0,1.5)
\LinAxis(0,200)(300,200)(4,5,-5,0,1.5)
\LogAxis(0,0)(0,200)(5,-5,0,1.5)
\LogAxis(300,0)(300,200)(5,5,0,1.5)
\Text(0,-10)[]{$100$}
\Text(75,-10)[]{$150$}
\Text(150,-10)[]{$200$}
\Text(225,-10)[]{$250$}
\Text(300,-10)[]{$300$}
\Text(140,-25)[]{$m_{A^0}$(GeV)}
\Text(-15,0)[]{$10^{-4}$}
\Text(-15,40)[]{$10^{-3}$}
\Text(-15,80)[]{$10^{-2}$}
\Text(-15,120)[]{$10^{-1}$}
\Text(-15,160)[]{$10^{0}~~$}
\Text(-15,200)[]{$10^{+1}$}
\Text(-40,100)[]{$\sigma$(fb)}
\Text(285,185)[]{$(a)$}
\Text(210,185)[]{$h^0$}
\Curve{(0.00E+00,165.4008)
(15.00000,167.7162)(30.00000,169.5210)
(45.00000,170.8860)(60.00000,171.8679)
(75.00000,172.5661)(90.00000,173.0975)
(105.0000,173.4922)(120.0000,173.7777)
(135.0000,174.0087)(150.0000,174.1832)
(165.0000,174.3450)(180.0000,174.4345)
(195.0000,174.5204)(210.0000,174.6336)
(225.0000,174.6974)(240.0000,174.6910)
(255.0000,174.7737)(270.0000,174.7967)
(285.0000,174.8183)(300.0000,174.8698)
}
\Text(210,165)[]{$h^0$}
\DashCurve{(0.00E+00,120.4000)
(15.00000,151.8389)(30.00000,169.2784)
(45.00000,170.7265)(60.00000,170.9796)
(75.00000,171.0371)(90.00000,171.0904)
(105.0000,171.1180)(120.0000,171.1192)
(135.0000,171.1410)(150.0000,171.1418)
(165.0000,171.1659)(180.0000,171.1415)
(195.0000,171.1445)(210.0000,171.1875)
(225.0000,171.1919)(240.0000,171.1319)
(255.0000,171.1711)(270.0000,171.1570)
(285.0000,171.1438)(300.0000,171.1668)
}{5}
\Text(210,120)[]{$H^0$}
\Curve{(0.00E+00,160.3631)
(15.00000,156.6521)(30.00000,152.5240)
(45.00000,148.2407)(60.00000,143.9445)
(75.00000,139.7392)(90.00000,135.7720)
(105.0000,132.0049)(120.0000,128.4506)
(135.0000,125.1613)(150.0000,122.0209)
(165.0000,119.1153)(180.0000,116.2878)
(195.0000,113.7364)(210.0000,111.3153)
(225.0000,109.0329)(240.0000,106.8381)
(255.0000,104.7880)(270.0000,102.8501)
(285.0000,101.0478)(300.0000,99.33869)
}
\Text(75,90)[]{$H^0$}
\DashCurve{(0.00E+00,170.1821)
(15.00000,164.2336)(30.00000,131.5618)
(45.00000,107.1183)(60.00000,91.96037)
(75.00000,81.09864)(90.00000,72.70076)
(105.0000,65.81241)(120.0000,59.95844)
(135.0000,54.92443)(150.0000,50.42268)
(165.0000,46.42968)(180.0000,42.72060)
(195.0000,39.43749)(210.0000,36.40994)
(225.0000,33.60905)(240.0000,30.96729)
(255.0000,28.53342)(270.0000,26.25905)
(285.0000,24.16072)(300.0000,22.19258)
}{5}
\Text(75,125)[]{$A^0$}
\Curve{(0.00E+00,121.4630)
(15.00000,120.5460)(30.00000,119.6741)
(45.00000,118.7393)(60.00000,117.8423)
(75.00000,116.9391)(90.00000,116.0957)
(105.0000,115.2007)(120.0000,114.3434)
(135.0000,113.5130)(150.0000,112.7069)
(165.0000,111.9056)(180.0000,111.0091)
(195.0000,110.2076)(210.0000,109.4341)
(225.0000,108.6008)(240.0000,107.7768)
(255.0000,106.9867)(270.0000,106.1727)
(285.0000,105.4003)(300.0000,104.6000)
}
\Text(75,45)[]{$A^0$}
\DashCurve{(0.00E+00,41.46299)
(15.00000,40.54601)(30.00000,39.67411)
(45.00000,38.73934)(60.00000,37.84234)
(75.00000,36.93906)(90.00000,36.09573)
(105.0000,35.20074)(120.0000,34.33522)
(135.0000,33.49174)(150.0000,32.67331)
(165.0000,31.86014)(180.0000,30.95241)
(195.0000,30.14006)(210.0000,29.35626)
(225.0000,28.51302)(240.0000,27.67942)
(255.0000,26.88011)(270.0000,26.05724)
(285.0000,25.27631)(300.0000,24.46763)
}{5}
%SM Higgs
\Text(210,150)[]{${\rm SM}~H$}
\DashCurve{(0.00E+00,174.8921)(15.00000,171.9342)
(21.60000,170.5771)
(30.00000,169.0104)
(43.34999,166.5208)
(45.00000,166.1339)
(60.00000,163.4503)(75.00000,160.9942)
(90.00000,158.4171)(105.0000,155.9984)
(120.0000,153.7187)(135.0000,151.4757)
(150.0000,149.4318)(165.0000,147.3734)
(180.0000,145.3609)(195.0000,143.3674)
(210.0000,141.6273)(225.0000,140.0318)
(240.0000,138.2339)(255.0000,136.6017)
(270.0000,135.1677)(285.0000,133.7131)
(300.0000,132.2148)
}{12}

%Panel (b)
\SetOffset(40,50)\SetWidth{1.}
\LinAxis(0,0)(300,0)(4,5,5,0,1.5)
\LinAxis(0,200)(300,200)(4,5,-5,0,1.5)
\LogAxis(0,0)(0,200)(5,-5,0,1.5)
\LogAxis(300,0)(300,200)(5,5,0,1.5)
\Text(0,-10)[]{$100$}
\Text(75,-10)[]{$150$}
\Text(150,-10)[]{$200$}
\Text(225,-10)[]{$250$}
\Text(300,-10)[]{$300$}
\Text(140,-25)[]{$m_{A^0}$(GeV)}
\Text(-15,0)[]{$10^{-4}$}
\Text(-15,40)[]{$10^{-3}$}
\Text(-15,80)[]{$10^{-2}$}
\Text(-15,120)[]{$10^{-1}$}
\Text(-15,160)[]{$10^{0}~~$}
\Text(-15,200)[]{$10^{+1}$}
\Text(-40,100)[]{$\sigma$(fb)}
\Text(285,185)[]{$(b)$}
\Text(140,-50)[]{Figure $5$}
\Text(15,95)[]{$h^0$}
\Curve{(0.00E+00,102.0330)
(15.00000,99.06134)(30.00000,96.11842)
(45.00000,93.37886)(60.00000,90.57100)
(75.00000,88.28439)(90.00000,86.04910)
(105.0000,84.23049)(120.0000,82.30526)
(135.0000,80.77772)(150.0000,79.68288)
(165.0000,78.55524)(180.0000,77.24454)
(195.0000,76.25438)(210.0000,75.56909)
(225.0000,74.72462)(240.0000,74.08914)
(255.0000,73.49717)(270.0000,72.78991)
(285.0000,72.62892)(300.0000,71.98789)
}
\Text(80,110)[]{$h^0$}
\DashCurve{(0.00E+00,183.7131)
(15.00000,176.8034)(30.00000,145.5705)
(45.00000,123.8117)(60.00000,111.1883)
(75.00000,102.9535)(90.00000,96.95046)
(105.0000,92.50976)(120.0000,88.80241)
(135.0000,85.85197)(150.0000,83.74512)
(165.0000,81.84586)(180.0000,79.91718)
(195.0000,78.34811)(210.0000,77.25533)
(225.0000,76.17532)(240.0000,75.13859)
(255.0000,74.32912)(270.0000,73.32515)
(285.0000,73.02650)(300.0000,72.18690)
}{8}
\Text(15,80)[]{$H^0$}
\Curve{(0.00E+00,83.51291)
(15.00000,86.29807)(30.00000,88.42359)
(45.00000,89.95159)(60.00000,90.78404)
(75.00000,91.06335)(90.00000,90.96338)
(105.0000,90.94767)(120.0000,90.31905)
(135.0000,89.80841)(150.0000,89.37515)
(165.0000,88.73939)(180.0000,87.87971)
(195.0000,87.20100)(210.0000,86.51979)
(225.0000,85.81318)(240.0000,84.97307)
(255.0000,84.04372)(270.0000,83.13002)
(285.0000,82.78670)(300.0000,81.72233)
}
\Text(15,150)[]{$H^0$}
\DashCurve{(0.00E+00,128.0120)
(15.00000,161.6464)(30.00000,179.2001)
(45.00000,179.7476)(60.00000,178.9452)
(75.00000,177.8770)(90.00000,176.6734)
(105.0000,175.8438)(120.0000,174.5335)
(135.0000,173.5001)(150.0000,172.5902)
(165.0000,171.6783)(180.0000,170.5233)
(195.0000,169.5845)(210.0000,168.6893)
(225.0000,167.7981)(240.0000,166.8364)
(255.0000,165.7327)(270.0000,164.7346)
(285.0000,164.2475)(300.0000,163.1187)
}{4}
\Text(15,110)[]{$A^0$}
\Curve{(0.00E+00,104.4009)
(15.00000,103.0429)(30.00000,101.8030)
(45.00000,100.4349)(60.00000,99.51701)
(75.00000,98.14738)(90.00000,96.99148)
(105.0000,95.96751)(120.0000,94.90418)
(135.0000,93.82291)(150.0000,92.59392)
(165.0000,91.76620)(180.0000,90.32925)
(195.0000,89.75514)(210.0000,88.58460)
(225.0000,87.79094)(240.0000,86.61140)
(255.0000,85.69642)(270.0000,84.68613)
(285.0000,83.99719)(300.0000,83.02842)
}
\Text(80,185)[]{$A^0$}
\DashCurve{(0.00E+00,184.4009)
(15.00000,183.0429)(30.00000,181.8030)
(45.00000,180.4349)(60.00000,179.5170)
(75.00000,178.1474)(90.00000,176.9915)
(105.0000,175.9675)(120.0000,174.9124)
(135.0000,173.8441)(150.0000,172.6275)
(165.0000,171.8116)(180.0000,170.3860)
(195.0000,169.8226)(210.0000,168.6624)
(225.0000,167.8787)(240.0000,166.7088)
(255.0000,165.8030)(270.0000,164.8016)
(285.0000,164.1212)(300.0000,163.1608)
}{5}

\end{picture}\\
\end{center}

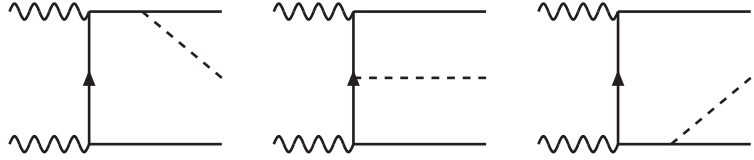
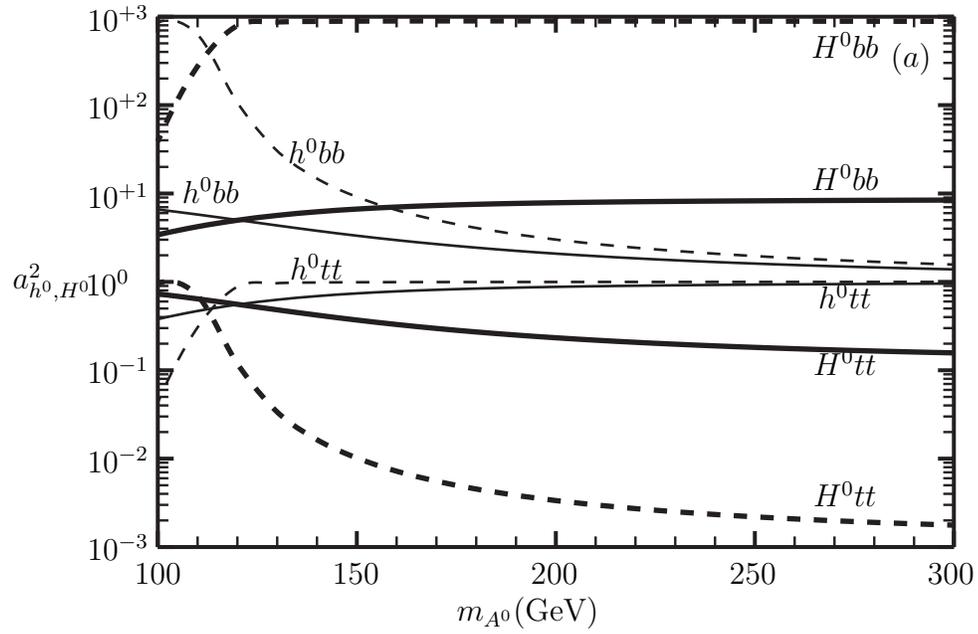
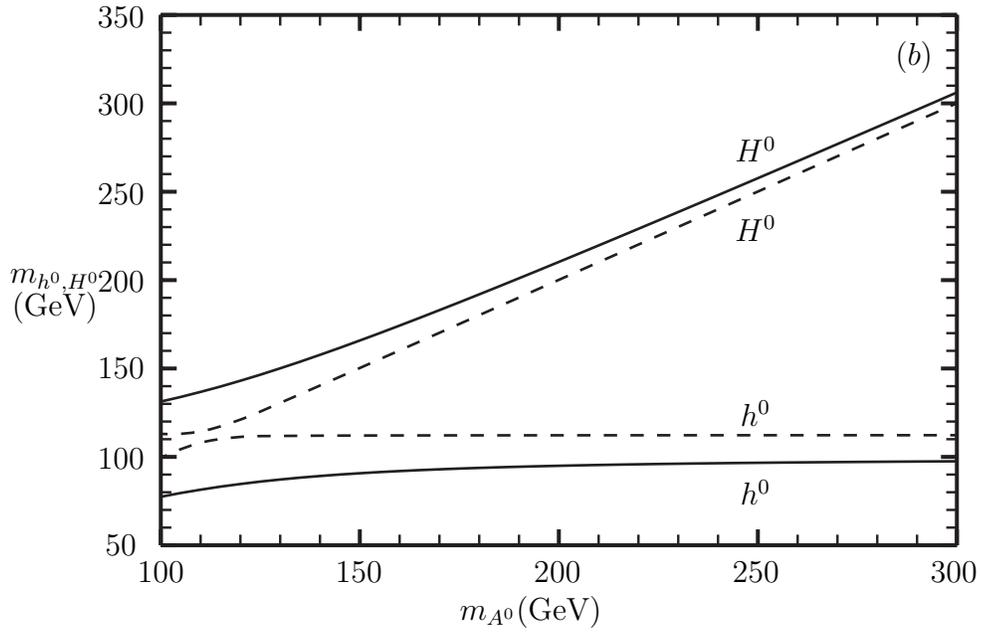
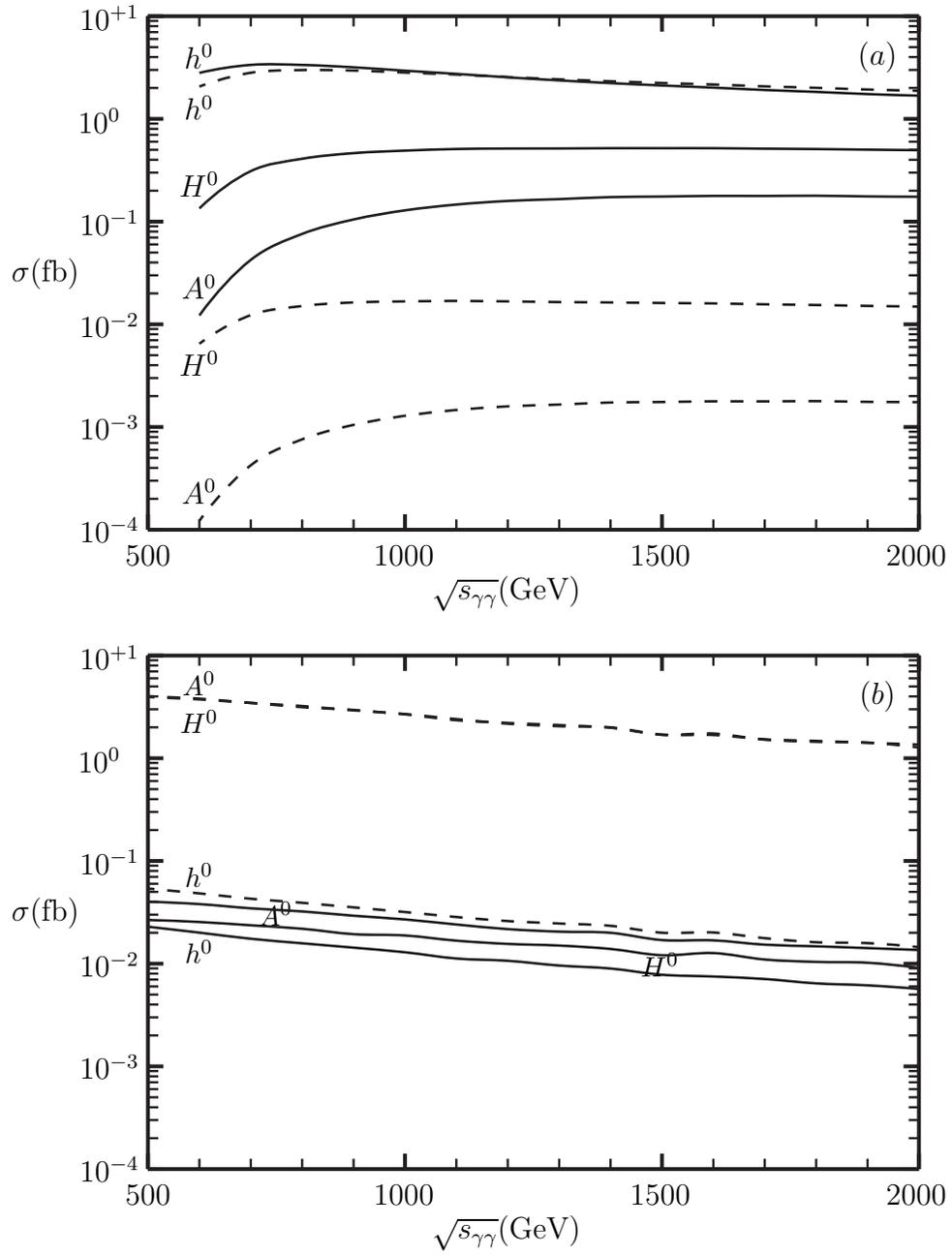
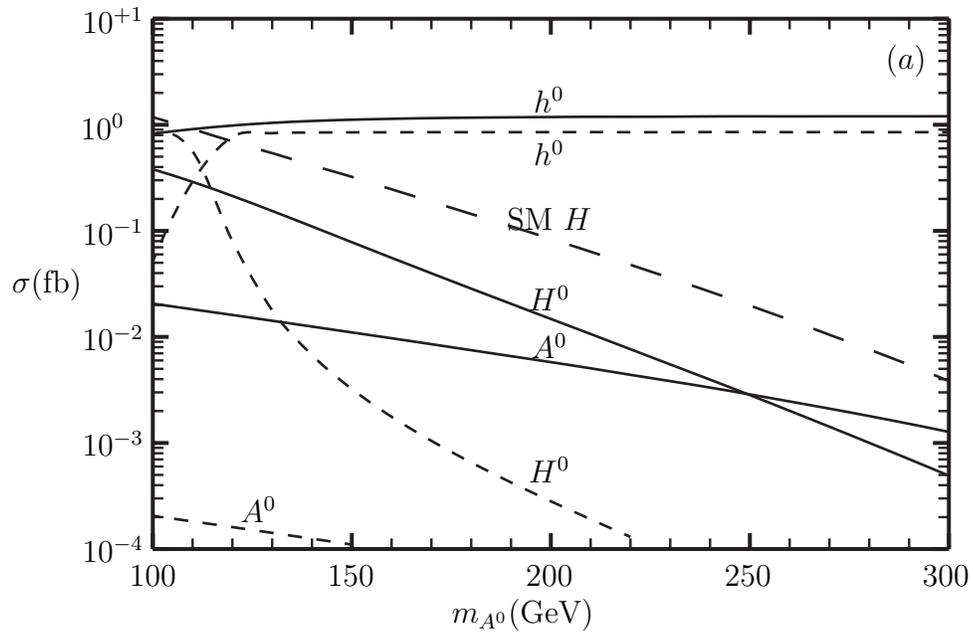
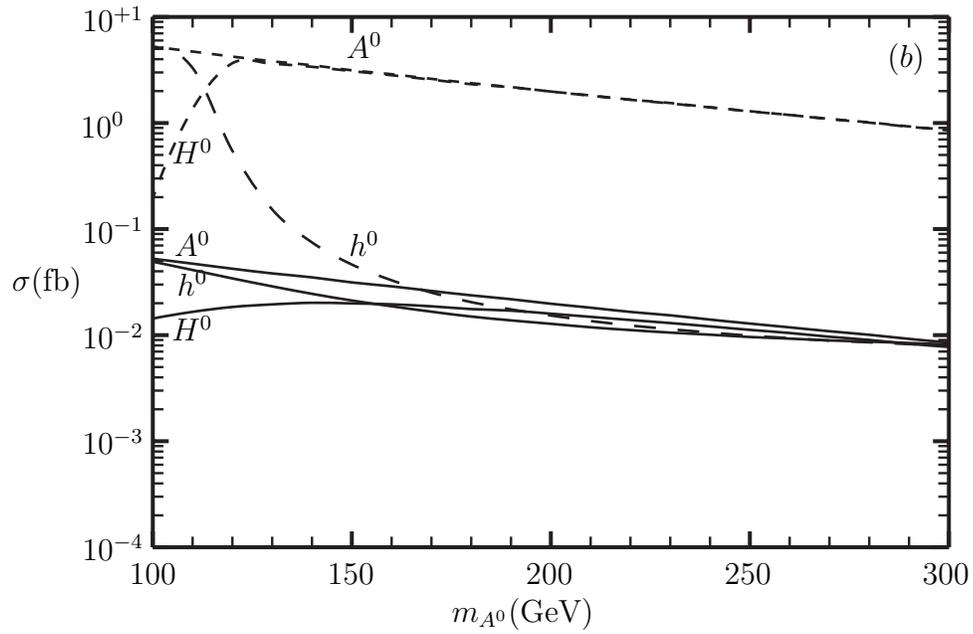
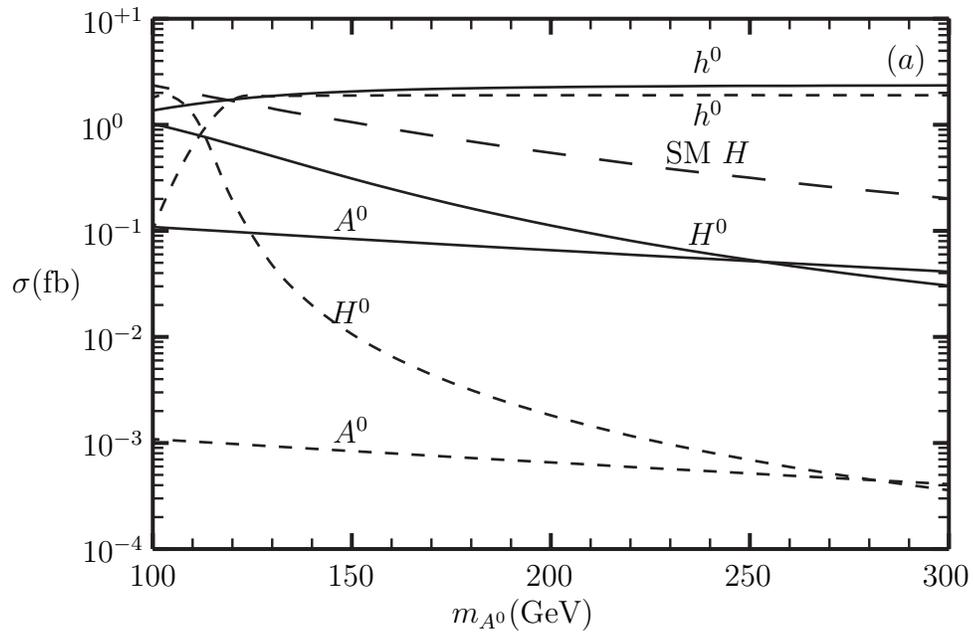
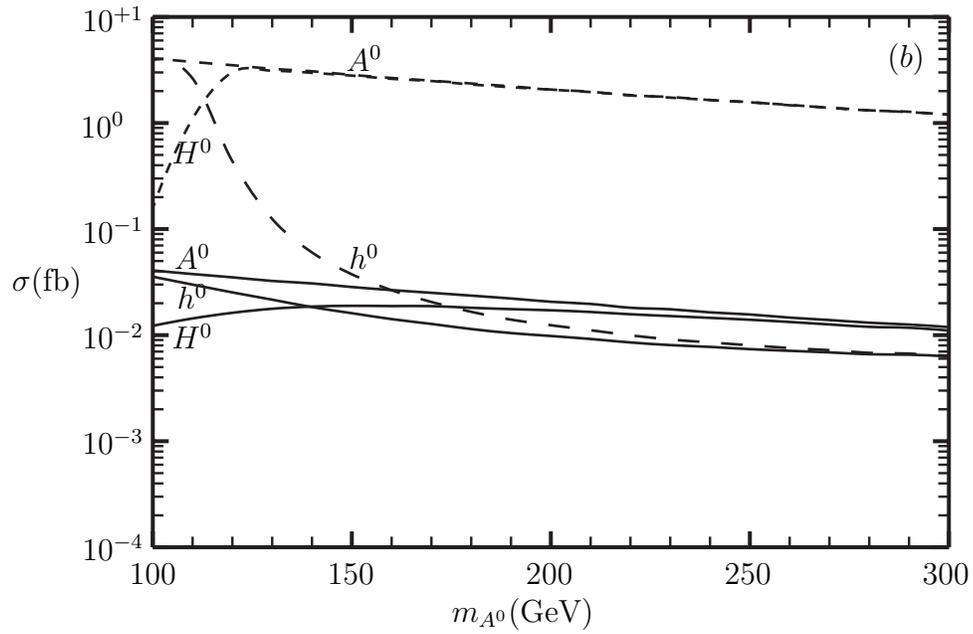
\end{document}